\documentclass[final,12pt]{elsarticle}

\usepackage{amsmath,amssymb,amsfonts}
\usepackage{lipsum}
\usepackage{hhline} % This package is for single cell but multi-column
\usepackage{graphicx} %Adding picture and figure
\usepackage{booktabs}
\usepackage{array}
\usepackage{url}
\usepackage{tabularx}
\usepackage{float}
\usepackage[numbers]{natbib}
\usepackage{subcaption}
\usepackage{hyperref}
\usepackage{tikz}
\usepackage{url}
\usetikzlibrary{shapes.geometric, arrows, positioning}

\tikzstyle{block} = [rectangle, draw, fill=blue!20, text width=6em, text centered, rounded corners, minimum height=3em]
\tikzstyle{data} = [rectangle, draw, fill=green!20, text width=6em, text centered, rounded corners, minimum height=3em]
\tikzstyle{comm} = [rectangle, draw, fill=orange!20, text width=8em, text centered, rounded corners, minimum height=3em]
\tikzstyle{arrow} = [thick,->,>=stealth]
\usepackage[justification=raggedright,singlelinecheck=false]{caption}

\begin{document}

\begin{frontmatter}



\title{An Energy-Efficient Smart Bus Transport Management System with Blind-Spot Collision Detection Ability} %% Article title

%% Author names and emails
\author[1]{Md. Sadman Haque}
\ead{mhaque221592@bscse.uiu.ac.bd}

\author[1]{Zobaer Ibn Razzaque}
\ead{zrazzaque221135@bscse.uiu.ac.bd}

\author[1]{Robiul Awoul Robin}
\ead{rrobin221564@bscse.uiu.ac.bd}

\author[1]{Fahim Hafiz\corref{cor1}}
\ead{fahimhafiz@cse.uiu.ac.bd}

\author[1]{Riasat Azim}
\ead{riasat@cse.uiu.ac.bd}

%% Corresponding author footnote
\cortext[cor1]{Corresponding author}

%% Affiliations
\affiliation[1]{organization={Department of Computer Science and Engineering, United International University},
            addressline={United City, Madani Avenue, Badda}, 
            city={Dhaka},
            postcode={1212}, 
            country={Bangladesh}}

%% Abstract
\begin{abstract}
Public bus transport systems in developing countries often suffers from a lack of real-time location updates and for users, making commuting inconvenient and unreliable for passengers. Furthermore, stopping at undesired locations rather than designated bus stops creates safety risks and contributes to roadblocks, often causing traffic congestion. Additionally, issues such as blind spots along with lack of following traffic laws, increase the chances of accidents. In this work, we address these challenges by proposing a smart public bus system along with intelligent bus stops that enhance safety, efficiency, and sustainability. Our approach includes a deep learning-based blind-spot warning system to help drivers avoid accidents with automated bus-stop detection to accurately identify bus stops, improving transit efficiency. We also introduce IoT-based solar-powered smart bus stops that show real-time passenger counts, along with an RFID-based card system to track where passengers board and exit. A smart door system ensures safer and more organised boarding, while real-time bus tracking keeps passengers informed. To connect all these features, we use an HTTP-based server for seamless communication between the interconnected network systems. Our proposed system demonstrated approximately 99\% efficiency in real-time blind spot detection while stopping precisely at the bus stops. Furthermore, the server showed real-time location updates both to the users and at the bus stops, enhancing commuting efficiency. The proposed energy-efficient bus stop demonstrated 12.71kWh energy saving, promoting sustainable architecture. Full implementation and source code are available at: \url{https://github.com/sadman-adib/MoveMe-IoT}
\end{abstract}

% 	\centering
% \includegraphics[width=1.2\textwidth]{Graphical_abstract.png}


%% Keywords

\end{frontmatter}



\section{Introduction}
\label{introduction}
%% Labels are used to cross-reference an item using \ref command.

In the present world, road traffic accidents are major global concerns, causing approximately 1.35 million deaths in 2016 according WHO \cite{lit28}. In 2017, WHO reported that around 1.3 million people died in road accidents, with 20-50 million people suffering non-fatal injuries that often lead to disabilities \cite{lit1}. In the USA, each year, around 60,000 buses are involved in accidents, leading to 14,000 and 300 non-fatal injuries and fatalities \cite{lit4}.\\
In \cite{lit11}, the study investigates the challenges faced by bus drivers in New Delhi, India, when stopping close to designated bus stops, highlighting factors such as passengers waiting on the roadway, insufficient bus handling capacity, and standing vehicles in the curb lane, which lead to improper stopping, traffic delays, and safety concerns. Similarly, \cite{lit14} examines the influence of bus stops on traffic flow in mixed traffic conditions in developing countries, emphasizing issues caused by buses not stopping properly at designated stops, resulting in traffic disruptions and increased congestion. In \cite{lit13}, the research explores how the design and placement of bus stops impact traffic flow and bus operations on suburban arterial roads, addressing the consequences of buses not adhering to designated locations, which affect traffic efficiency and safety. Complementing these findings, \cite{lit12} discusses the impact of bus stop location and design on passenger safety and traffic flow, emphasizing the problems of buses not stopping at designated stops, leading to increased waiting times and traffic congestion. It creates unsafe conditions for both other vehicles and passengers on the road. These unpredictable stopping patterns can lead to congestion and increase the risk of accidents. After that, passengers waiting at bus stops often face uncertainty regarding seat availability on incoming buses. Passengers may struggle to decide whether to board a bus or wait for the next one, or avail any other kind of transportation, leading to crowdy and inefficiencies in the transportation system. Again traditional bus-stops are not designed with energy efficiency. Most of the bus-stops lack sustainable power sources.\cite{lit16} explores using an off-grid solar power system to electrify bus stops in Melaka, Malaysia, with a 500-watt setup to support sustainable public transport. Similarly, \cite{lit15} examines the potential of solar energy for bus shelters in Ávila, Spain, using LiDAR to study shadow impacts. Both studies show how solar power can improve bus stop functionality and support eco-friendly urban growth. So still in many countries, the absence of renewable energy solutions in bus-stops, such as solar-powered lighting and digital displays, results in less informative waiting areas, reducing convenience and safety for passengers, and at the time is night.

To address these aforementioned challenges in public bus transportation system in developing countries, we propose a smart bus system that integrates advanced features for improved sustainability, safety, efficiency. Our framework includes machine learning CNN (YOLO) based light-weight object detection for blind-spot object and bus-stop detection, solar-powered bus-stops, a smart bus-door system, an RFID based passenger counting system that displays real-time count data for incoming buses at bus-stop and a software that will track the location of the bus along with a software for the bus owners to track the total count of passengers on hourly, weekly or monthly basis.\\

\section{Literature Review}
\label{Lit_review}
%% Labels are used to cross-reference an item using \ref command.

This section reviews related studies on smart and energy-efficient transportation systems, focusing on how machine learning, the Internet of Things, and renewable energy sources are integrated to improve efficiency and performance. These studies highlight how combining these technologies can help reduce energy use, lower costs, and make transportation systems more sustainable. By using advanced technologies, these systems aim to create safer, more efficient, and eco-friendly transportation solutions.

In \cite{lit1}, a vision-based blind-spot warning (BSW) system using DNN was proposed, integrating YOLOv3 and Detectron2 for vehicle detection, Dense Depth for depth estimation and OpenCV for visualization. Another real-time, camera-based BSW system with self-supervised deep learning was introduced in \cite{lit2}, with 0.97 accuracy achievement with a light-weight neural model and pre-trained object detector for distance estimation and tracking vehicles. In \cite{lit3}, blind-spot detection in autonomous vehicles has also been explored, where a lightweight camera based system using a FCN (Fully Connected Network) was trained on vehicle and non-vehicle datasets. Combining custom designed CNNs with ResNet-50 and ResNet-101 architecture, a multi-CNN-based blind-spot detection system was developed for vehicles in \cite{lit4}. In \cite{lit10}, a cost-effective blind-spot detection system on TinyML was presented, leveraging lightweight models. The system achieved 0.88 accuracy. Table~\ref{comT} presents an overview of the some relevant blind-spot detection methods used, including the datasets, hardware, and associated limitations.

\begin{table}[htbp]%% placement specifier
\caption{Recent works on blind spot detection.}
% \small
\scriptsize
	\begin{tabular}{|p{1.5cm}|p{1.5cm}|p{1.7cm}|p{2cm}|p{2cm}|p{3cm}|}%% Table column specifiers
		\hline %% Horizontal line at the top
		%% Tabular cells are separated by &
		\textbf{Ref.} & \textbf{Method Used} & \textbf{Datasets} & \textbf{Hardware} & \textbf{Detection Performance} & \textbf{Limitations} \\ %% A tabular row ends with \\
		\hline %% Horizontal line at the top
		\cite{lit1} & DNN and Depth Estimation & KITTI datasets & Tesla P100 (16GB) GPU & below 50\% & The system lacks retraining on KITTI, limiting accuracy, and needs testing in diverse real-world conditions \\
		\hline %% Horizontal line at the top
		\cite{lit2} & Self-Supervised Deep Learning Model & KITTI, COCO, Own Dataset & Not mentioned & 97\% & This study lacks a generic way to classify dangerous situations, sufficient training data, efficient data collection, and context awareness for reliable detection.\\
		\hline %% Horizontal line at the top
		\cite{lit3} & Deep Learning & Own Dataset & N1s4, 4vCPU, 15GB, 200GB & 93.75\% & This study lacks sufficient image datasets, live video testing, diverse weather evaluations, guaranteed performance, and deep learning-based comparison analysis. \\
		\hline %% Horizontal line at the top
		\cite{lit4} & Deep CNN Architecture &  LISA Dataset, Own Dataset & Xeon E-2124G, 32GB RAM , Quadro P4000 & 98\% & The study is limited by computational cost, dataset constraints, lower frame rates, restricted CNN integration, and applicability mainly to heavy vehicles. \\
		\hline %% Horizontal line at the top
		\cite{lit10} & Lite Object Detection Algorithm & Pre-train Datasets & Arduino Nano & 88.24\% & The study is limited by a small dataset, untested performance in varied conditions, and processing constraints due to low-cost hardware.\\
		\hline %% Horizontal line at the top
	\end{tabular}
	%% Use \caption command for table caption and label.
    \label{comT}
\end{table}
In \cite{lit5}, renewable energy integration into public transport has been explored, where photovoltaic systems were integrated into trolleybus networks using Monte Carlo simulations to assess energy efficiency. A similar study in \cite{lit6}, analyzed solar-powered electric buses in university campuses, evaluating different solar energy setups. The research analyzed that a PV system with a 375 kW capacity in open spaces was the most cost-effective and it ensured reduced fossil fuel consumption and energy efficiency. An IOT-based public transport management system has been investigated in \cite{lit7}, where a system utilizing multiple sensors, an Arduino Uno, and a cloud server was implemented for bus tracking and real time data collection. In \cite{lit8}, a broader review of smart transport systems technologies was conducted with IOT, ML, 4G/5G, V2V and V2X. It covers some features like route optimization, accident detection while addressing challenges like scalability and data privacy. Smart bus-stops utilizing IOT for energy efficiency and real time monitoring were proposed in \cite{lit9}. Reduced energy consumption and improved fault detection, with solar power integration for sustainability, a prototype was demonstrated.
All the existing research focuses on specific aspects of public transportation, whereas our study integrates multiple advanced features into a framework, creating a novel solution that will revolutionize the public bus transportation system.

\section{Methodology}
\label{method}
\begin{figure}[H]
    \centering
    \includegraphics[width=1\textwidth]{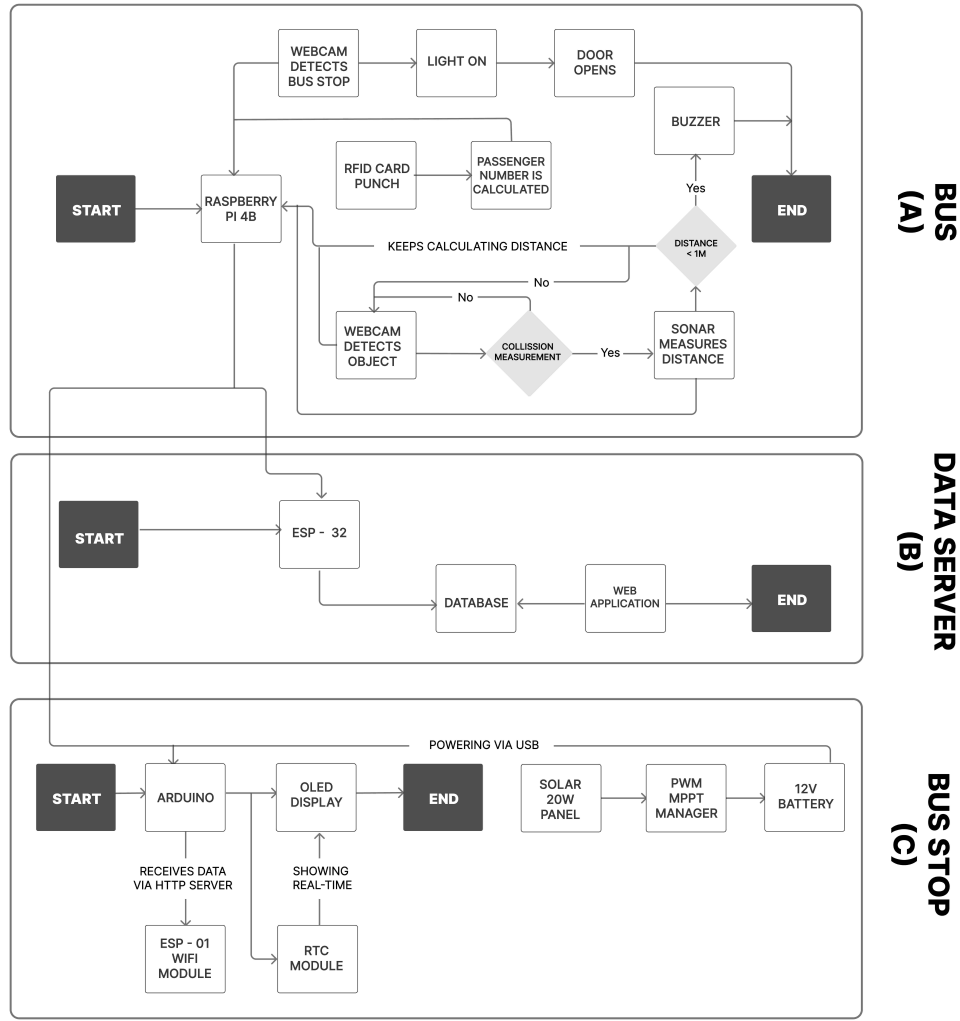} % Adjust width as needed
    \caption{Full system architecture. It illustrates a smart bus monitoring and collision detection system divided into three sections: BUS (A), DATA SERVER (B), and BUS STOP (C). In the BUS (A) section, the process starts with a Raspberry Pi 4B, where a webcam detects bus stops, turning on the light and opening the door. Passengers use RFID card punch, and the system calculates the passenger count. The Raspberry Pi continuously measures distance; if a webcam detects an object, collision measurement is triggered, and a sonar sensor measures the distance. If the distance is less than 1 meter, a buzzer sounds as a warning. The DATA SERVER (B) section starts with an ESP32, which transmits data to a database, making it accessible via a web application. In the BUS STOP (C) section, an Arduino receives real-time data via an ESP-01 WiFi module and an RTC module, displaying it on an OLED screen. The system is powered through a USB connection from a solar 20W panel, which charges a 12V battery via a PWM MPPT manager. This architecture integrates real-time bus tracking, collision alerts, and passenger updates for an efficient transportation system.}
    \label{sa}
\end{figure}

\begin{figure}[H]
    \centering
    \includegraphics[width=1\textwidth]{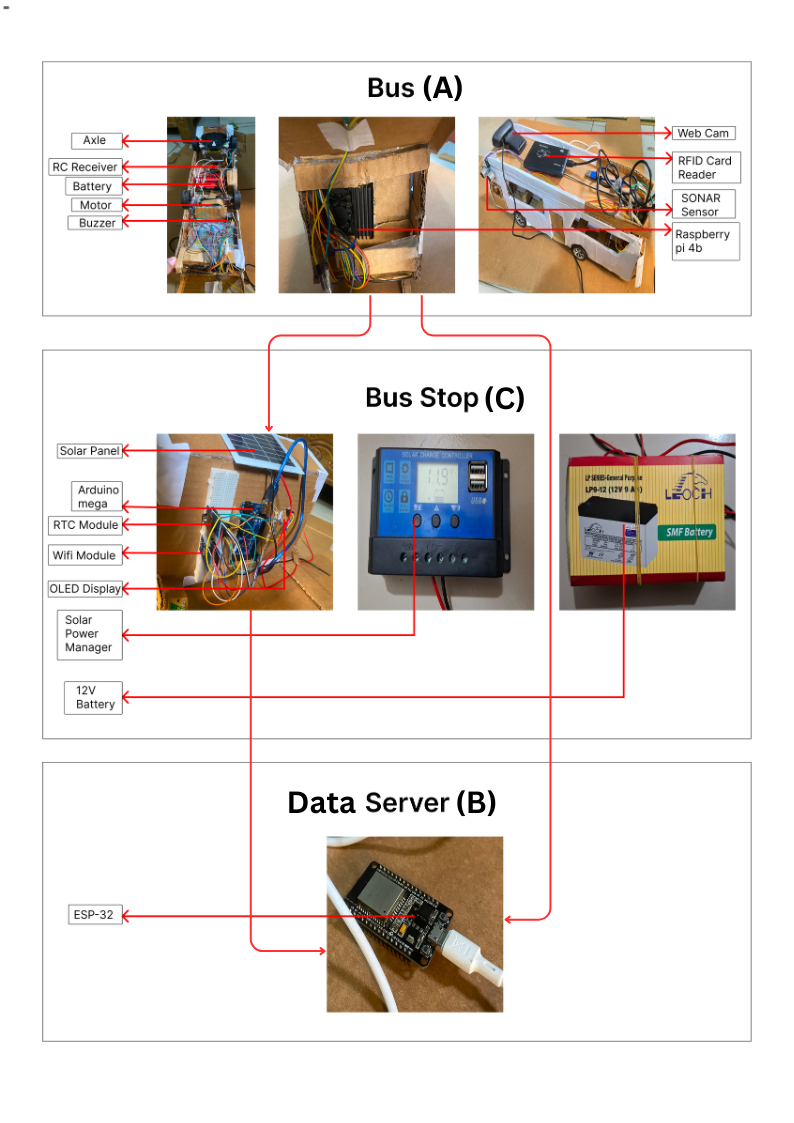} % Adjust width as needed
    \caption{Fully detailed implementation architecture. It explains the hardware architecture, showcasing the integration of a Raspberry Pi 4B with camera and sonar sensors for object detection, an LED for bus stop alerts, an RFID system for passenger tracking, and wireless networking for real-time seat count updates to bus-stop displays, all powered by onboard electricity and solar energy.}
    \label{fig: detail}
\end{figure}

\begin{figure}[H]
	\centering
	\includegraphics[width=1\textwidth]{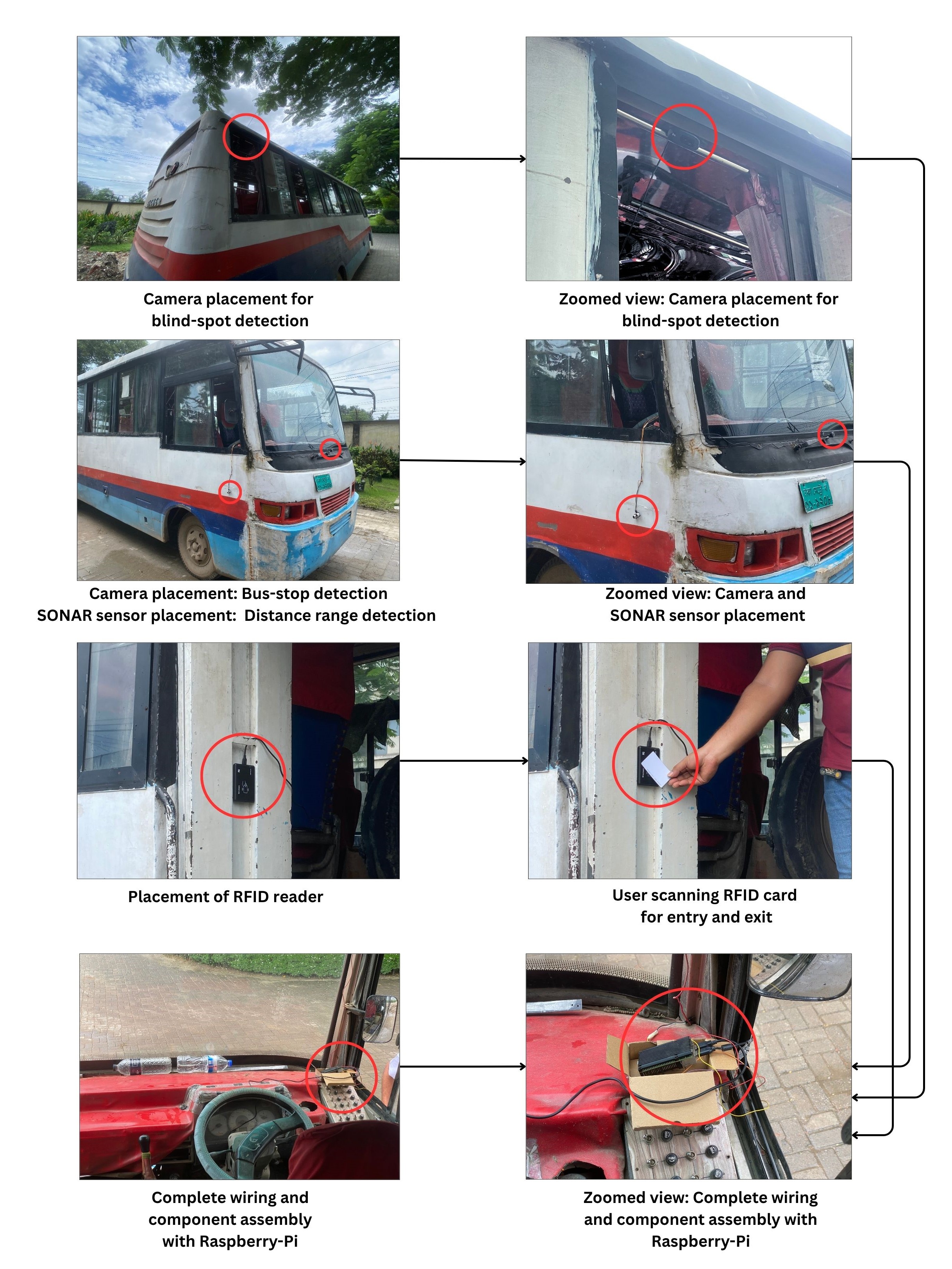} % Adjust width as needed
	\caption{Component setup in bus for real-time implementation. It shows the complete setup of a smart bus monitoring system, highlighting the practical placement of cameras for blind-spot and bus-stop detection, SONAR sensors for distance measurement, an RFID reader for logging passenger entry and exit, and the integration of all components with a Raspberry Pi inside the bus. The images illustrate how each device is mounted and connected, ensuring real-time safety monitoring, obstacle detection, and automated passenger tracking within a single, compact system.}
	\label{fig: Implemented_real_time}
\end{figure}

%% Labels are used to cross-reference an item using \ref command.
Our proposed and implemented system integrates ML-driven object detection inside blind-spot and warning, using camera and sonar sensors to detect nearby objects and trigger buzzer alerts if a collision risk is detected, meaning a detected object is within the 1 meter proximity of the vehicle. Also, the camera will detect bus stops, activating an LED light to alert the driver. The Smart Bus-Door System ensures that the door opens only at the bus stop unless it is in an emergency. An RFID based passenger tracking system updates real-time seat count on bus-stop displays via wireless networking.The Raspberry Pi 4b is powered by the electricity available onboard. Bus-stop display is solar powered, ensuring energy efficiency. A web application provides admins with real-time data, and seat count information, which is password protected and is hosted on the cloud. The full system architecture is presented in Figure~\ref{sa}, the detailed hardware integration of the Raspberry Pi, sensors, RFID, and wireless modules is shown in Figure~\ref{fig: detail}, and the complete setup of a smart bus monitoring system is shown in Figure~\ref{fig: Implemented_real_time}. The overall bus architecture is divided into three blocks- Block A, Block B, and Block C, as demonstrated in Figure~\ref{sa} and Figure~\ref{fig: detail}. The following sections briefly discuss these blocks from Section~\ref{subsec:blockA} to \ref{subsec:blockC}.

\subsection{Proposed system architecture of the BUS (Block A)}
\label{subsec:blockA}

The overall hardware system implemented in the Bus can be divided again into three parts, which are the blind-spot detection system, the bus-stop detection system, and the smart bus-door system. These sub-building blocks are discussed from Subsection~\ref{subsubsec:blind-bus} to \ref{subsubsec:smart_door}.
\subsubsection{Blind-spot collision detection System}
\label{subsubsec:blind-bus}
\begin{figure}[H]
    \centering
    \includegraphics[width=0.8\textwidth]{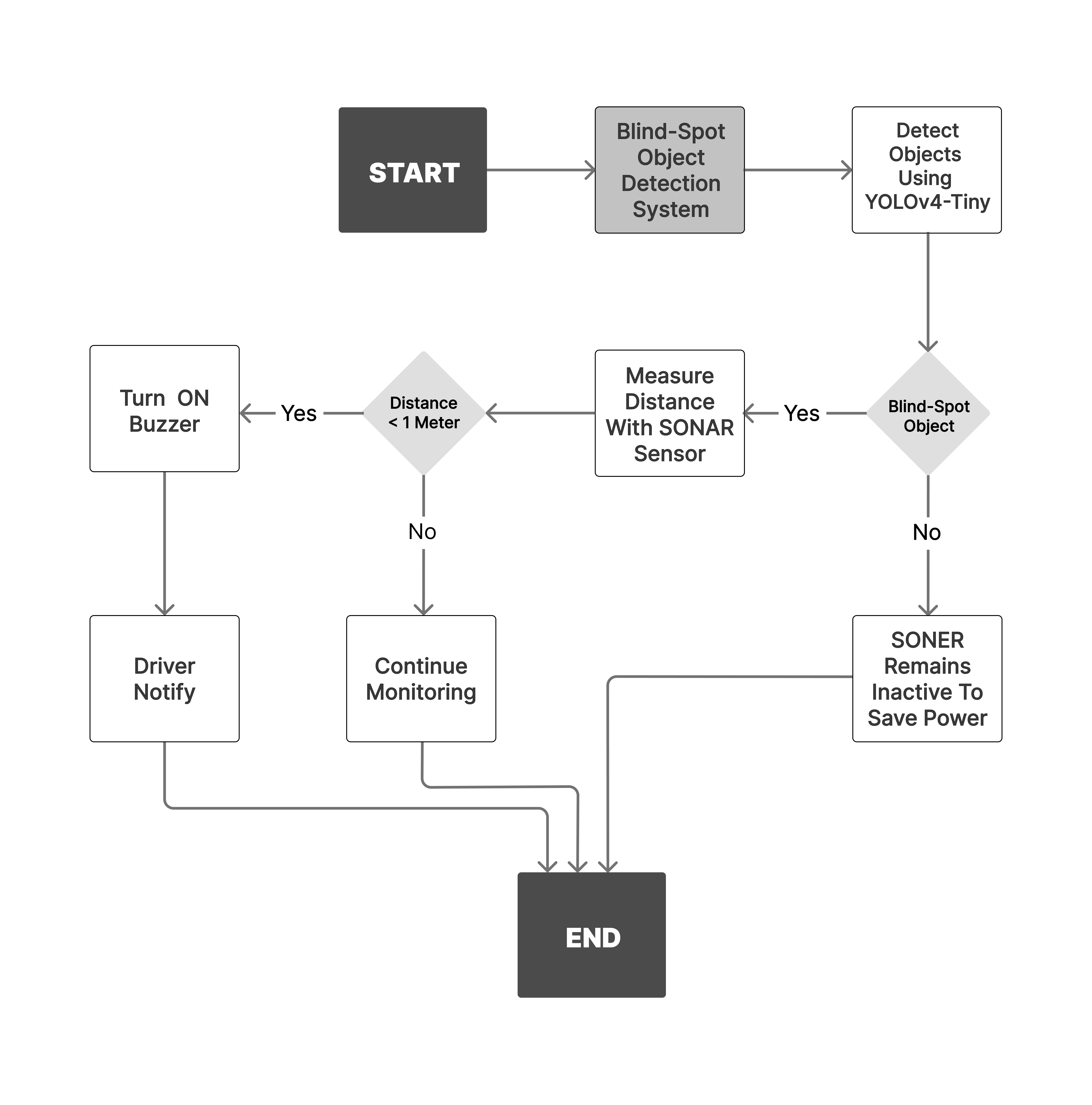} % Adjust width as needed
    \caption{Blind-spot object detection architecture. It shows that the blind-spot detection system uses a YOLOv4-Tiny model on a Raspberry Pi 4b to detect objects around the bus, activates a SONAR sensor for distance measurement, and triggers a buzzer alarm if a collision risk is detected within 1m, enhancing safety.}
    \label{blindspot Arch}
\end{figure}

The blind-spot object detection system uses a CNN based machine learning model (YOLOv4-Tiny) running on a Raspberry Pi 4b to identify objects around the buses, such as persons, trucks, buses, cars, motorbikes etc, through a camera mounted on the body of the bus. When an object is detected, the system activates the ultrasonic (sonar) sensor to measure the distance of the objects from the bus. If no object is detected, the sensor remains inactive to save power. If an object is detected and the SONAR detects its presence within a critical distance (less than 1m), indicating a possible collision, the system turns on a buzzer alarm to immediately notify the driver. This real-time processing enhances safety by providing instant feedback, reducing blind-spot-related accidents. The blind-spot detection system with YOLOv4-Tiny, SONAR sensor, and buzzer for collision alerts is shown in Figure~\ref{blindspot Arch}.

\subsubsection{Bus-stop detection System}
\label{subsubsec:bus_stop}
The bus-stop detection system utilizes a camera mounted with the bus, connected to a Raspberry Pi 4b running an ML model (YOLOv4-Tiny) to detect bus-stops. For the prototype we use the stop sign, which is already trained in datasets in YOLOv4-Tiny as bus-stops. When a bus-stop is detected, the Raspberry Pi processes the data and turns on an LED light at the front of the area of driver. The LED light serves as a visual alert, indicating to the driver that he is approaching a bus-stop and should prepare to stop. This system enhances driver awareness and improves safety by providing timely, automated notification based on real-time video processing.

\subsubsection{Smart Bus-Door System System}
\label{subsubsec:smart_door}
\begin{figure}[H]
	\centering
	\includegraphics[width=0.8\textwidth]{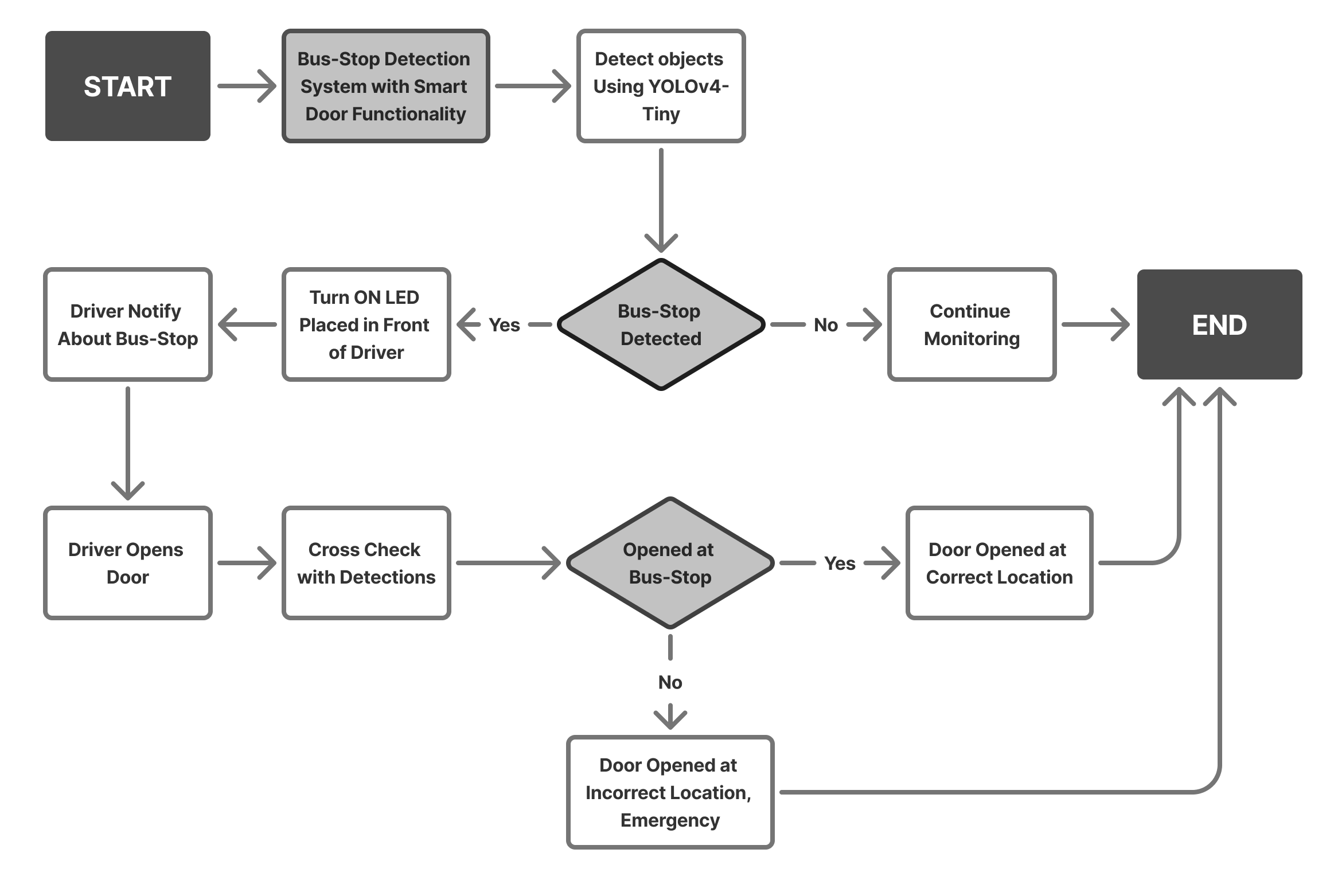} % Adjust width as needed
	\caption{Bus-stop detection system with smart bus-door functionality architecture. The system employs a YOLOv4-Tiny model on a Raspberry Pi 4b to detect bus-stops via a camera, activating an LED to alert the driver for increased safety and awareness. Simultaneously, a smart bus-door mechanism cross-checks the detected location with actual bus-stops to ensure doors only open at valid stops, issuing warnings and triggering emergency alerts if opened incorrectly, thus enhancing both safety and operational efficiency.}
	\label{fig: Smart Bus-Door System Architecture}
\end{figure}

The smart bus-door system is designed to ensure that bus doors open at bus-stops for passengers safety and operational efficiency. Now after successfully detecting the bus-stop, when the driver opens the door, the system cross-check the current location with the detected bus-stop. The system provides an alert confirming that the door was opened in the current location of the bus-stop. If the door is opened away from the bus-stop, the system automatically triggers an emergency alert that the driver has opened the door in an inappropriate location in an emergency situation. This smart system helps prevent unauthorized door openings and enhances safety by providing real-time feedback on the operational states of buses. The smart bus-door system, which cross-checks detected locations with actual bus-stops to ensure doors open only at valid stops and provides alerts to enhance safety and operational efficiency, is shown in Figure~\ref{fig: Smart Bus-Door System Architecture}.

\subsection{Proposed Data Server/Network System (Block B)}
\label{subsec:blockB}
\begin{figure}[H]
	\centering
	\includegraphics[width=0.8\textwidth]{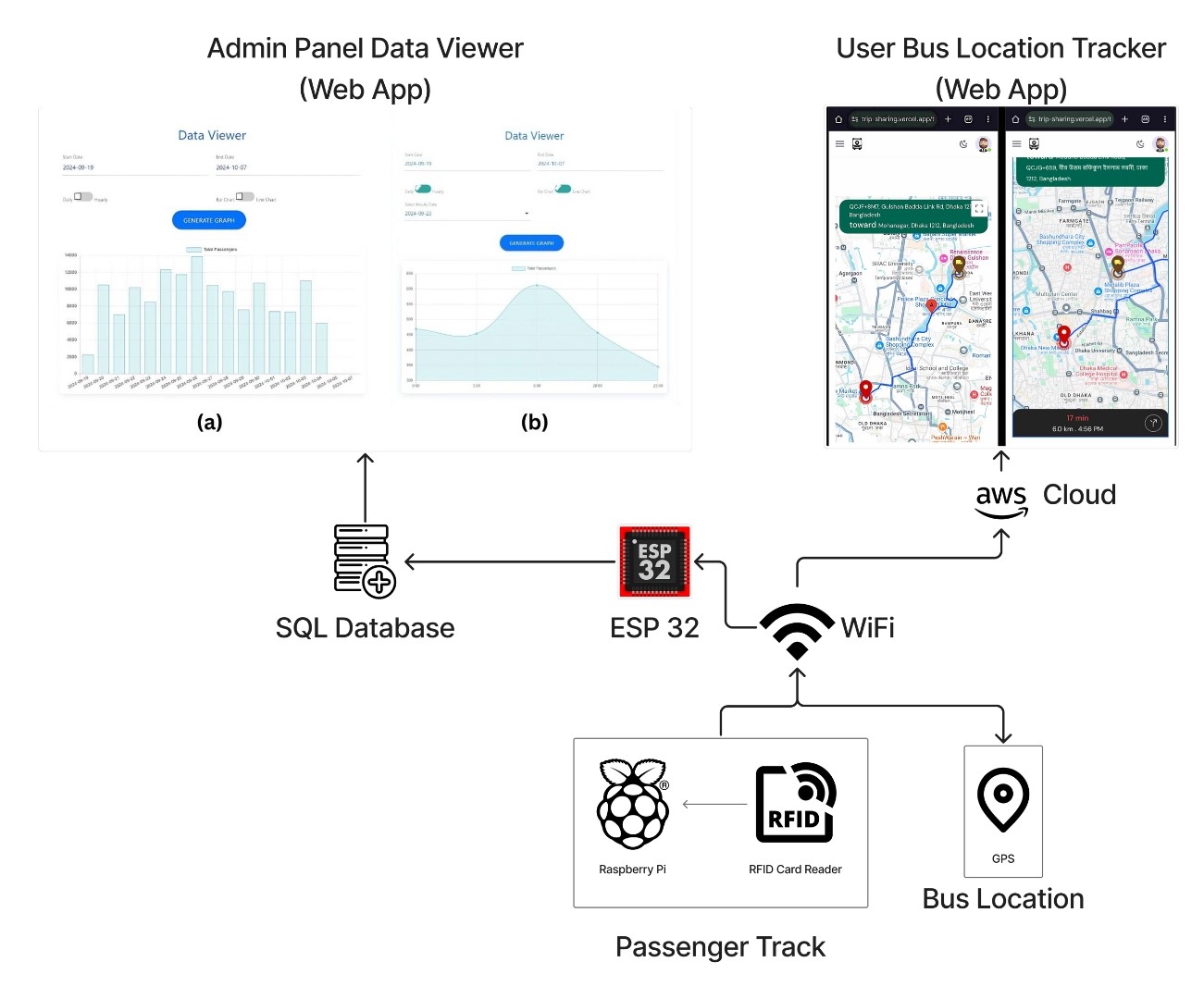} % Adjust width as needed
	\caption{Smart transportation interface. This figure illustrates the data visualization components of the system, including the passenger data viewer, which uses bar and line charts to display passenger counts and trends over time with interactive filtering options, and the traffic visualization interface, which provides real-time route mapping with traffic conditions, estimated travel time, distance, and key location markers to support navigation and route optimization.}
	\label{fig:side_by_side}
\end{figure}

To ensure a seamless experience and to provide a smooth visual experience to the users and admins, we have created two separate web applications, one is done in React.js front-end, Node.js back-end and utilized Google Maps API key to pinpoint the location of the bus and to provide real time updates of the road traffic condition. Using this app, users can have a proper view on how far the next bus is, how long it will take for the bus to reach and the condition of the road. The second app is for the bus admins, they can login and figure out the frequency of the passengers using the bus. The system that utilizes RFID and GPS to track passengers and bus locations, with data transmitted via ESP32 to the cloud and visualized through admin and user web applications for real-time monitoring and analysis, is shown in Figure~\ref{fig:side_by_side}.

\subsection{Proposed Solar powered Bus-stop System (Block C)}
\label{subsec:blockC}
\begin{figure}[H]
	\centering
	\includegraphics[width=0.8\textwidth]{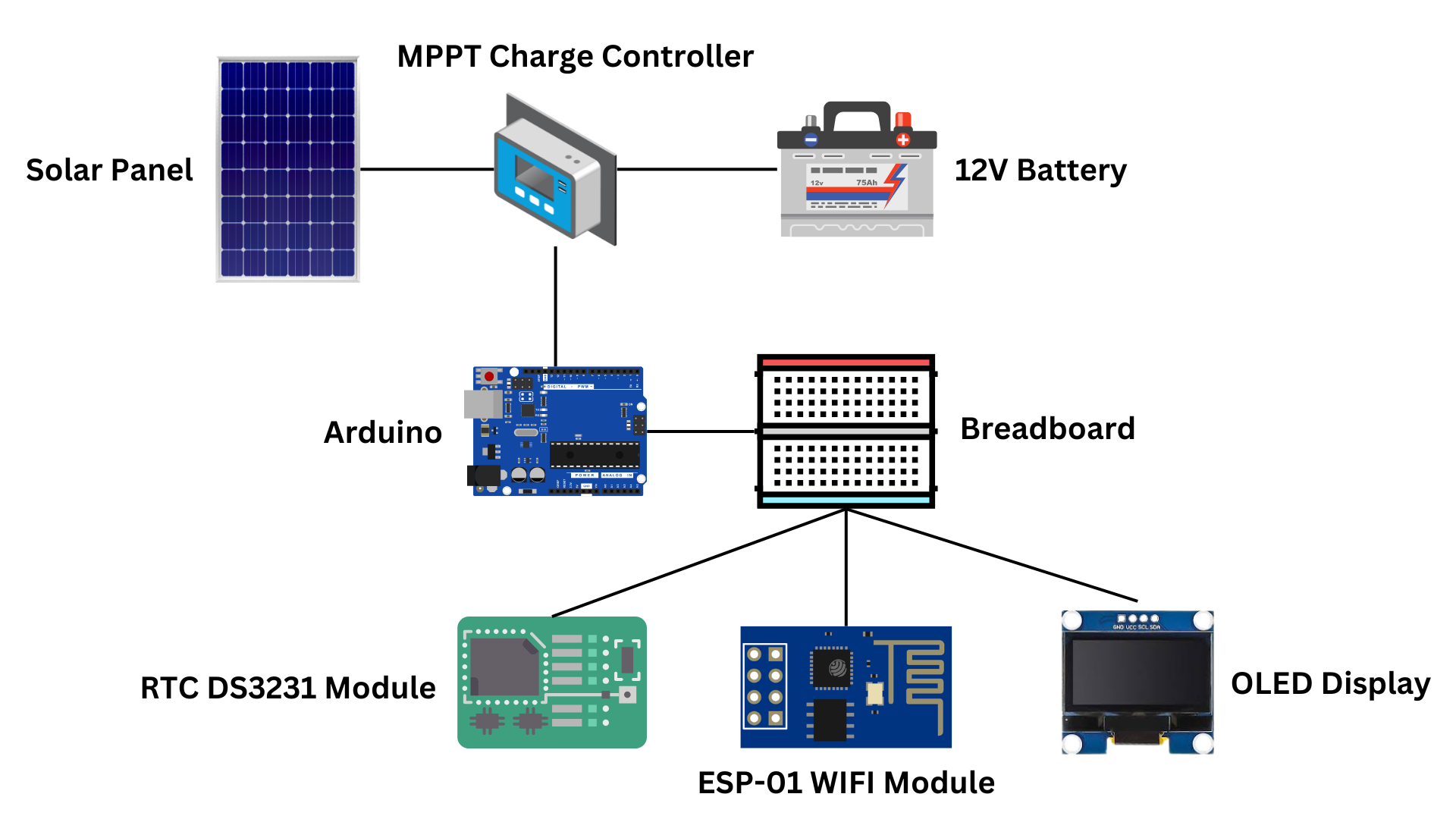} % Adjust width as needed
	\caption{Data visualization diagram of Block C (Bus-Stop). It shows the Visualization Diagram, illustrating the integration of a solar-powered system with an MPPT charge controller, a 12V battery, and an Arduino microcontroller. The setup includes a breadboard for circuit connections, an RTC DS3231 module for real-time clock functionality, an ESP-01 WiFi module for wireless communication, and an OLED display for real-time data visualization, enabling efficient data visualization and management.}
	\label{fig: circuit diagram}
\end{figure}

\begin{figure}[H]
	\centering
	\includegraphics[width=0.8\textwidth]{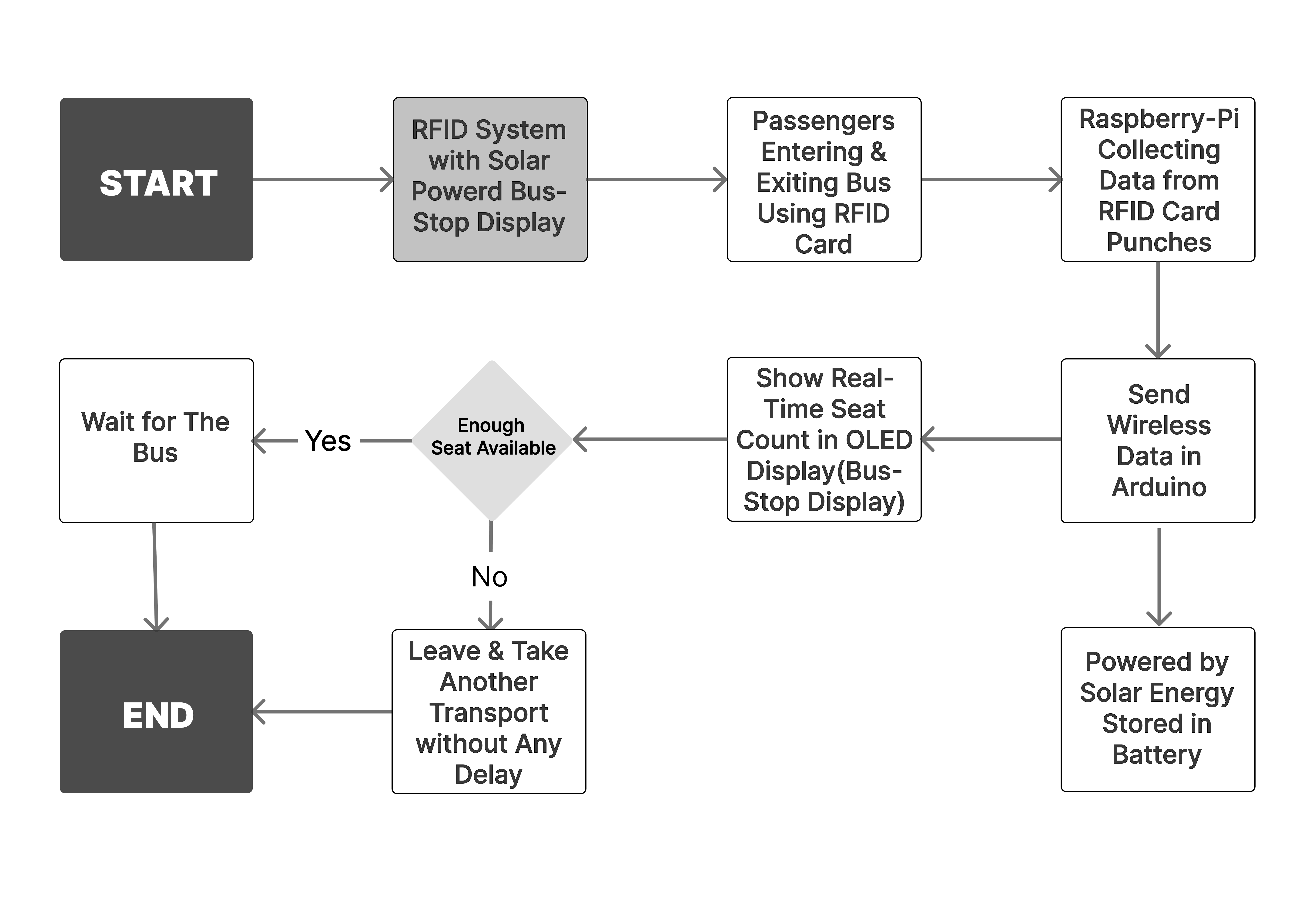} % Adjust width as needed
	\caption{RFID data count and visualization system architecture. It shows an RFID-based passenger tracking system, where a Raspberry Pi collects entry and exit data from RFID cards on the bus and wirelessly transmits it via socket programming to an Arduino-powered bus-stop display, which, running on solar energy, provides real-time seat availability to assist passengers in planning their commute efficiently. }
	\label{fig: RFID System Architecture}
\end{figure}

The RFID system is installed on the bus to track when passengers get on and off from the bus. The Raspberry Pi collects this data by reading RFID cards as passengers enter and exit. This information is sent wireless via socket programming to display at the bus-stop, which is powered by arduino. The display shows real-time seat count information on the bus, helping passengers know the bus status. To make the system energy efficient, the bus-stop display is powered by solar energy stored in batteries, so it can keep running even without sun-light. This system makes it easier for passengers to know when a bus is arriving and how many seats are available in that bus. The data visualization diagram of Block C (Bus-Stop), illustrating the solar-powered system integration with MPPT, Arduino, and real-time data display, is shown in Figure~\ref{fig: circuit diagram}; the RFID-based passenger tracking system for real-time seat availability is shown in Figure~\ref{fig: RFID System Architecture}.

%% Use \subsubsection, \paragraph, \subparagraph commands to 
%% start 3rd, 4th and 5th level sections.
%% Refer following link for more details.
%% https://en.wikibooks.org/wiki/LaTeX/Document_Structure#Sectioning_commands

%% Refer following link for more details.
%% https://en.wikibooks.org/wiki/LaTeX/Mathematics
%% https://en.wikibooks.org/wiki/LaTeX/Advanced_Mathematics

%% Use a table environment to create tables.
%% Refer following link for more details.
%% https://en.wikibooks.org/wiki/LaTeX/Tables

%% Use figure environment to create figures
%% Refer following link for more details.
%% https://en.wikibooks.org/wiki/LaTeX/Floats,_Figures_and_Captions

\section{Results}
\label{result}
To evaluate our proposed IoT-based smart bus management system's performance, we discuss the results in two folds: i) First, the accuracy of the object detection capability of our system in real-time is discussed in Section~\ref{result:Blind-spot} and Section~\ref{result:energy_eff} discusses the energy efficiency of our proposed system in detail.

\subsection{Blind-spot Object Detection Results} 
\label{result:Blind-spot}

To analyze the detection results of blind spot objects, this study considers several key evaluation parameters, including True Positives (TP), True Negatives (TN), False Positives (FP), False Negatives (FN), Precision, Recall, F1 Score, False Discovery Rate (FDR), and the Confusion Matrix. Evaluation metrics like precision and recall rely directly on TP, TN, FP, and FN values \cite{lit18}. Another study highlights the advantages of the Matthews Correlation Coefficient (MCC) over the F1 Score, as MCC uses all four parameters more evenly to provide a balanced measure \cite{lit17}. The relationship between MCC and other metrics shows that MCC can approach the geometric mean of precision and recall as TP, TN, FP, and FN vary, particularly as TN increases \cite{lit19}. In this study, True Positives were recorded when the system correctly detected vehicles or pedestrians in the blind spot and issued appropriate warnings. True Negatives occurred when no objects were present and the system did not produce unnecessary alerts. False Positives arose when background elements were mistakenly identified as obstacles, causing unwarranted alerts, while False Negatives were noted when the system failed to detect an actual vehicle or person in the blind spot, which could compromise safety. As highlighted by \cite{lit22}, precision, recall, and F1 score are essential for evaluating classification models, especially with imbalanced datasets, as precision reduces false positives, recall minimizes false negatives, and the F1 score balances both for robust detection performance.\\

\textbf{True Positives (TP):} This happens when the system correctly detects an object that is actually there. It means the system successfully identifies something it should detect. Ensures real hazards or relevant objects are recognized, improving safety and reliability.  \\

\textbf{True Negatives (TN):}  This is when the system correctly identifies that there is nothing to detect. It means the system understands that no object is present and avoids false alerts. Prevents false alarms and builds user or driver trust. \\

\textbf{False Positives (FP):} This occurs when the system mistakenly thinks it sees an object that is not actually there. It is like a false alarm where the system detects something that does not exist. Can lead to unnecessary actions, hesitation, or reduced confidence in the system.  \\

\textbf{False Negatives (FN):} This happens when the system fails to notice an object that is really there. In this case, the system misses something important that it should have detected. This can be highly dangerous, potentially leading to collisions, overlooked hazards, or serious safety issues.\\

\textbf{Precision:} When false positives are costly, Precision is useful. Precision shows how accurate the positive predictions of the model are by calculating the percentage of correct positive results out of all positive predictions made. It helps us understand how often the model is right when it says something is positive \cite{lit18}, \cite{lit20}.
\begin{equation}
	Precision = \frac{TP}{TP + FP}
\end{equation}

\textbf{Recall:} When missing positives are costly, Recall is important. Recall measures how well the model finds all the actual positive cases. It tells us how good the model is at not missing any positive instances \cite{lit19}, \cite{lit21}.
\begin{equation}
	Recall = \frac{TP}{TP + FN}
\end{equation}

\textbf{FDR:} FDR (False Discovery Rate) is important when controlling FP is crucial. FDR shows how often the positive predictions of the model are wrong. It calculates the percentage of false positives out of all positive predictions, helping us see how many mistakes the model makes when identifying positives \cite{lit21}, \cite{lit19}.
\begin{equation}
	False Discovery Rate (FDR) = 1 - \text{Precision}
\end{equation}

\textbf{F1 Score:} F1 score balances precision and recall. Where both FP and FN are problematic, F1 score is important. It is useful when the data is imbalanced because it considers both how accurate and how complete the positive predictions of the model are \cite{lit17}, \cite{lit18}.
\begin{equation}
	F1 Score = \frac{2 \cdot (\text{Precision} \cdot \text{Recall})}{\text{Precision} + \text{Recall}}
\end{equation}

\begin{figure}[H]
	\centering
	\includegraphics[width=1\textwidth]{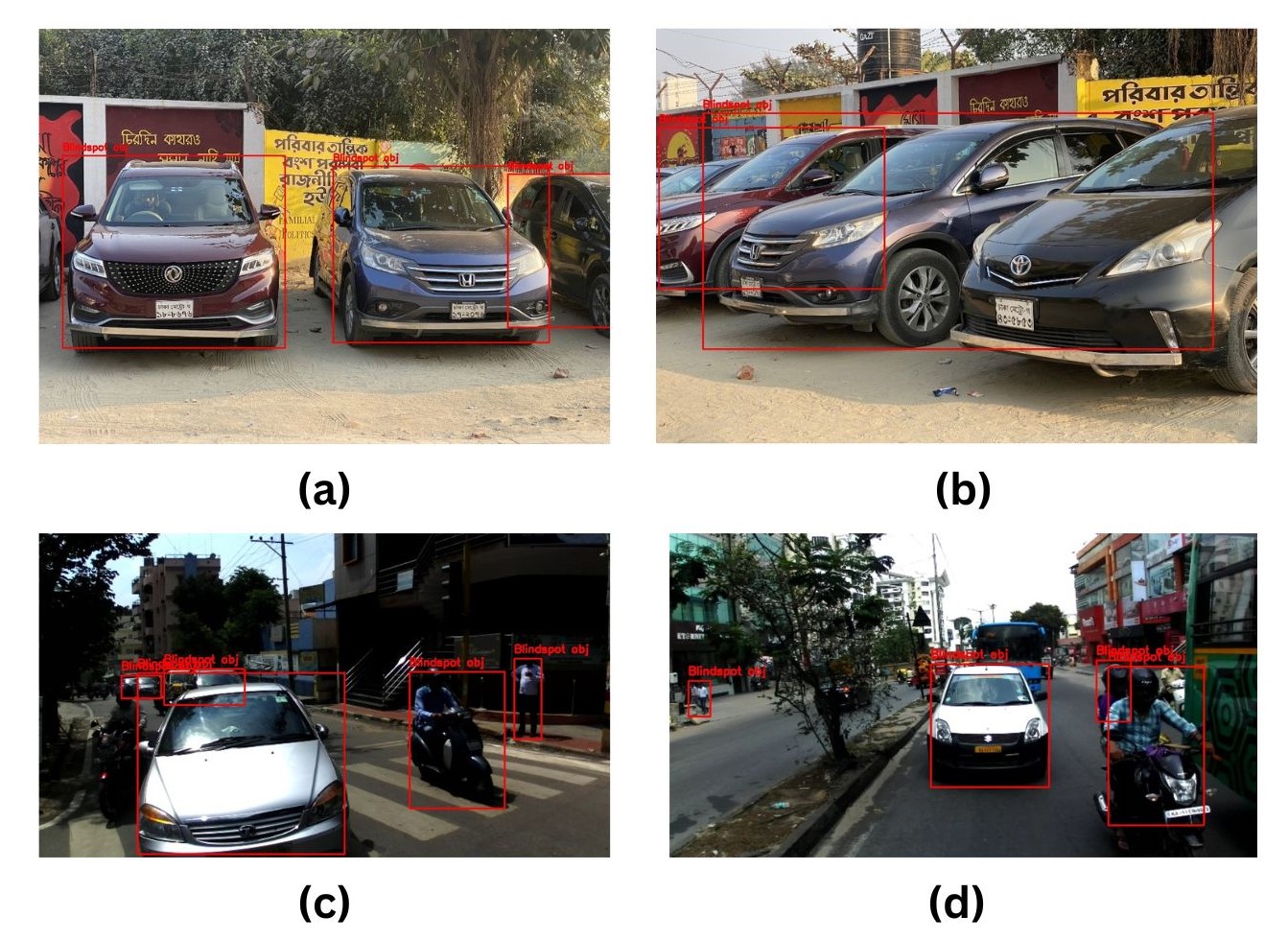} % Adjust width as needed
	
	\caption{Blind-spot collision detection sample frames. It (a, b, c, d) illustrates all objects detected within the blind spots of the bus, highlighting them with distinct red bounding boxes}
	\label{fig:four_images}
\end{figure}

In our case, True Positives were recorded when the system accurately detected vehicles or pedestrians in the blind spot, alerting the driver appropriately. True Negatives were observed when no objects were present, and the system did not issue any unnecessary warnings. However, False Positives occurred when background elements were mistakenly identified as obstacles, leading to unnecessary alerts. False Negatives were noted when the system failed to detect an actual vehicle or person in the blind spot, potentially compromising safety. During the blind spot object detection process, we monitored the road area for about 10 minutes, capturing many frames from the live camera feed. To evaluate the effectiveness of the system, we analyzed 10 different frames to see if it accurately detected potentially dangerous objects in the blind spot, like vehicles or pedestrians that the driver might not notice. This helped us determine whether the system could reliably identify harmful objects and alert the driver, enhancing safety by making them aware of hidden dangers. The blind-spot collision detection sample frames, highlighting detected objects within the bus’s blind spots, are shown in Figure~\ref{fig:four_images}; the blind-spot object detection results per frame, including true positives, false positives, false negatives, and true negatives, are presented in Table~\ref{frameanalysis}. The performance summary, including overall precision, recall, FDR, and F1 score, is presented in Table~\ref{performanceanalysis}. \\

\begin{figure}[H]
	\centering
	\includegraphics[width=0.8\textwidth]{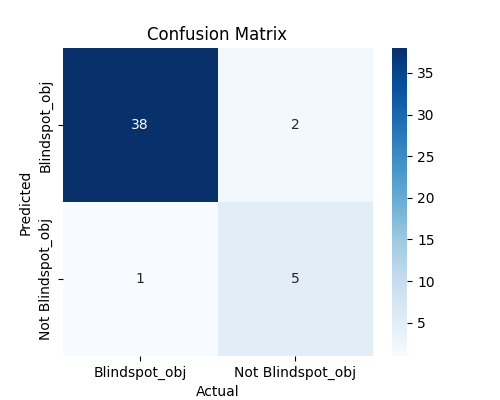} % Adjust width as needed
	\caption{Confusion matrix. It shows how well the blind-spot detection model is working. It correctly detected blind-spot objects 38 times and correctly ignored unnecessary objects 5 times. However, it made 2 mistakes by wrongly identifying an unnecessary object as a blind-spot object when it was not, and it missed 1 actual blind-spot object. Overall, the model is performing well, with most detection being correct and only a few errors. }
	\label{fig: Confusion Matrix}
\end{figure}

\begin{table}[H]%% placement specifier
	\caption{Blind-spot object detection per frame. The table demonstrates an example upto 10 real-time frames.}
\centering
\begin{tabular}{|l|p{3em}|p{3em}|p{3em}|p{3em}|p{12em}|}
\hline
\textbf{Frame no.} & \textbf{TP} & \textbf{FP} & \textbf{FN} & \textbf{TN} & \textbf{Interpretation} \\
\hline
Frame 1 & 8 & 1 & 0 & 1 & Detected 8 objects correctly, correctly ignored 1, but falsely detected 1 object. No misses. \\
\hline
Frame 2 & 3 & 0 & 0 & 0 & All 3 detections were correct. No negatives present to test TN or FP. \\
\hline
Frame 3 & 2 & 0 & 0 & 0 & Detected 2 objects correctly. No negatives to evaluate. \\
\hline
Frame 4 & 4 & 0 & 0 & 0 & All 4 detections were accurate. No negatives involved. \\
\hline
Frame 5 & 1 & 0 & 1 & 1 & Detected 1 object, missed 1 actual object, but correctly ignored 1. \\
\hline
Frame 6 & 2 & 0 & 0 & 0 & Clean detection of 2 objects. No negatives present. \\
\hline
Frame 7 & 6 & 0 & 1 & 0 & Detected 6 objects, but missed 1. No TN or FP. \\
\hline
Frame 8 & 5 & 0 & 0 & 0 & All 5 detections were correct. No negatives involved. \\
\hline
\end{tabular}
	%% Use \caption command for table caption and label.
\label{frameanalysis}
\end{table}
After calculating True Positives (TP), True Negatives (TN), False Positives (FP), and False Negatives (FN), we can organize these values into a Confusion Matrix. This matrix in Figure~\ref{fig: Confusion Matrix} shows the actual versus predicted classifications, helping us evaluate the performance of the model, including accuracy, precision, recall, FDR, F1 score \cite{lit20, lit21}.
% \textbf{Confusion Matrix:} A Confusion Matrix evaluates the performance of a classification model by comparing actual and predicted classifications, detailing True Positives, True Negatives, False Positives, and False Negatives to analyze accuracy and errors \cite{lit20}. It also visualizes the relationship between actual and predicted labels, helping to understand model effectiveness and classification errors across different classes \cite{lit21}. \\

\begin{table}[H]%% placement specifier
\caption{Performance table summary}
\centering%% For centre alignment of tabular.
\begin{tabular}{|l|c|}%% Table column specifiers
 \hline %% Horizontal line at the top
%% Tabular cells are separated by &
  \textbf{Type} & \textbf{Result} \\ %% A tabular row ends with \\
   \hline %% Horizontal line at the top
  Overall Precision & 98.8\% \\
   \hline %% Horizontal line at the top
  Overall Recall & 93.6\% \\
   \hline %% Horizontal line at the top
   Overall FDR & 1.2\% \\
   \hline %% Horizontal line at the top
  F1 Score & 96.1\% \\
   \hline %% Horizontal line at the top
\end{tabular}
%% Use \caption command for table caption and label.
\label{performanceanalysis}
\end{table}

\begin{figure}[H]
	\centering
	\includegraphics[width=1\textwidth]{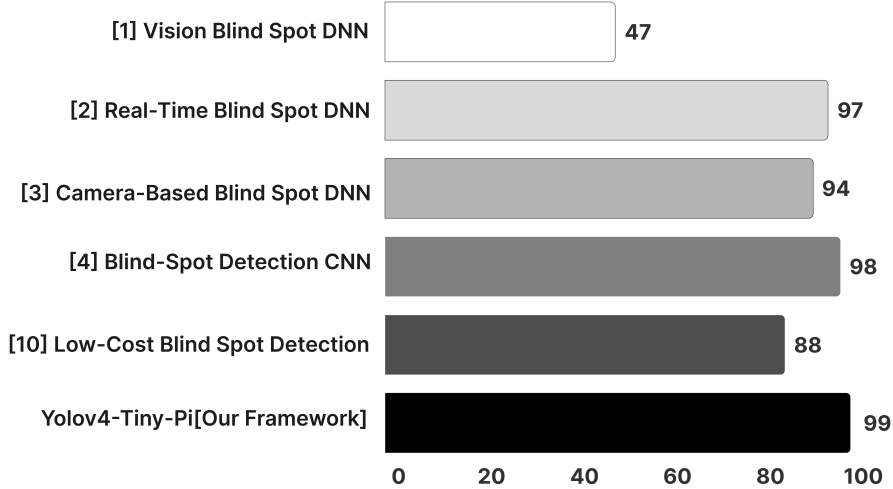} % Adjust width as needed
	\caption{Blind-spot object detection performance comparison with state-of-the-art models. It shows that the Yolov4-Tiny-Pi framework achieves a detection performance of 98.8\%, which is almost 99\%, matching the highest-performing models, while other methods range from below 50\% to 98\%, demonstrating its competitive accuracy.}
	\label{fig: sota}
\end{figure}

\noindent Our framework utilizes a CNN-based deep learning approach with COCO and pre-trained datasets (YOLOv4--Tiny), running on a Raspberry Pi 4B (4GB), achieving an overall precision of 98.8\%. This demonstrates shown in Figure~\ref{fig: sota} the capability of the Raspberry Pi 4 to handle high-performance detection tasks efficiently despite its constrained hardware resources. YOLOv4 is efficient due to its optimized architecture and feature enhancements. It integrates CSPDarkNet53, which reduces computation while maintaining high accuracy, making it suitable for real-time applications \cite{lit25}. The use of a Bidirectional Feature Pyramid Network (BiFPN) improves feature fusion, enhancing small object detection \cite{lit27}. Additionally, Scaled-YOLOv4 optimizes cross-stage partial (CSP) connections, increasing efficiency and adaptability across different hardware constraints \cite{lit26}. The improved YOLOv4-tiny further enhances computational efficiency by incorporating residual blocks and auxiliary networks, ensuring high-speed object detection with reduced complexity \cite{lit24}. \\

\subsection{Energy Efficiency With Solar System Results}
Our proposed system is energy efficient with a solar-powered system which have a significant impact considering today's energy demand in the world. This is a unique contribution in the literature as the previous works lack any integration of an energy-efficient approach in the Bus management system. We discuss the energy efficiency of our system by examining the energy consumption for a bus stop in Subsection~\ref{subsubsec_result:energy_consump}. We then calculate the required solar panel size for such bus stops in Subsection~\ref{subsubsec_result:solar_size}.
\label{result:energy_eff}.

\subsubsection{Energy Consumption Calculation for the Bus-stop}
\label{subsubsec_result:energy_consump}
Energy consumption is the amount of power used by a system to work. Using solar energy helps reduce this by converting sunlight into electricity, which powers things like buses and bus stops without relying on traditional power sources. This saves money and cuts down on pollution, making cities cleaner and more efficient.
Calculating energy consumption is essential for optimizing the integration of renewable energy in urban transportation systems, ensuring efficiency, sustainability, and reliability. In trolleybus systems, it helps determine the optimal size of photovoltaic systems to maximize solar energy utilization \cite{lit5}. For smart city bus stops, accurate energy consumption analysis enables efficient power management, enhancing energy efficiency and operational sustainability \cite{lit9}. Additionally, understanding energy needs is crucial in evaluating the feasibility of solar-powered bus shelters, ensuring realistic predictions of energy generation \cite{lit15}, \cite{lit16}. According to \cite{lit23}, energy efficiency with solar energy is necessary to enable continuous autonomous operation of IoT devices without requiring regular maintenance or power infrastructure access. The daily energy consumption of components, including the I2C OLED display, Arduino Mega, WiFi module, RTC module, and LED light, is presented in Table~\ref{tab:energy_consumption}.
% \textbf{Total Daily Energy Consumption (Wh):}

\begin{equation}
	EnergyConsumptionDaily = Power \times \text{Daily Usage}
\end{equation}

\begin{table}[H]
    \caption{Daily energy consumption of our proposed system's components}
    \centering
    \renewcommand{\arraystretch}{1.3} % Adjust row height for better readability
    \begin{tabular}{|m{2.5cm}|m{1.5cm}|m{1.5cm}|m{3cm}|m{3cm}|}
        \hline
        \textbf{Components} & \textbf{Power (W)} & \textbf{Daily Usage (Hours)} & \textbf{Min Energy Consumption (Daily)} & \textbf{Max Energy Consumption (Daily)} \\
        \hline
        0.96-inch I2C OLED Display & 0.02-0.15 & 24 & $0.02 \times 24 = 0.48$ & $0.15 \times 24 = 3.6$ \\
        \hline
        Arduino Mega (ATmega2560-based) & 0.1-0.5 & 24 & $0.1 \times 24 = 2.4$ & $0.5 \times 24 = 12$ \\
        \hline
        Arduino WiFi Module (ESP8266 or ESP32-based) & 0.02-0.7 & 24 & $0.02 \times 24 = 0.48$ & $0.7 \times 24 = 16.8$ \\
        \hline
        RTC Module (DS3231) & 0.0001-0.001 & 24 & $0.0001 \times 24 = 0.0024$ & $0.001 \times 24 = 0.024$ \\
        \hline
        5mm LED Light & 0.01-0.1 & 10 & $0.01 \times 10 = 0.1$ & $0.1 \times 10 = 1$ \\
        \hline
    \end{tabular}
    \label{tab:energy_consumption}
\end{table}

\noindent Total Minimum Energy Consumption = 3.46 Wh = 0.0035 kWh \\
Total Maximum Energy Consumption = 34.824 Wh = 0.0348 kWh.

\noindent Daily energy consumption is key to designing a solar power system. This ensures the system can generate and store enough energy to meet the demand throughout the day and night. By calculating total daily energy consumption we can size the solar panel and battery correctly. This ensures the system runs without interruption even on low sunlight days.\\

\subsubsection{Solar Panel Sizing for the proposed bus stop}
\label{subsubsec_result:solar_size}
Solar panel sizing is needed to ensure that the panels generate enough energy to meet the maximum consumption of the system while minimizing waste. By accurately sizing the panels based on energy needs, daily usage, and sunlight availability, we can optimize costs, maximize efficiency, and ensure reliable power supply.\\

\begin{equation}
	 RequiredPower_{\text{solar}} = \frac{EnergyConsumption}{AverageDailySunlightHours}
\end{equation}
Assuming 5 Peak Sunlight Hours per day:\\
Minimum: $\frac{3.46 \text{ Wh}}{5 \text{ h}}$ = 0.69 W (Negligible, so a small solar panel will work)\\
Maximum: $\frac{34.824 \text{ Wh}}{5 \text{ h}}$ = 6.964 W (7 W solar panel)\\
Recommended: A 10W solar panel should be more than enough to meet this demand.

To cover daily energy needs the solar panel must generate enough electricity to match the available sunlight hours. The solar panel is sized based on peak sunlight hours so it can generate enough energy within a short time frame. A slightly larger panel is recommended to account for weather conditions variation of sunlight intensity. The \textbf{Battery Backup Requirement} can be calculated as follows:\\
For night-time and cloudy days:\\
Minimum: 3.46 Wh = 12V, 0.29 Ah  (or 24V, 0.14Ah)\\
Maximum: 34.824 Wh = 12V, 2.78Ah (or 24V, 1.39Ah)\\
Recommended: A 12V 5Ah battery should be sufficient for extended backup.

Since solar panels only generate power during the day a battery is needed to store excess energy for night and cloudy days. The battery capacity is calculated based on energy consumption so there is enough backup power. A slightly larger battery is recommended to maintain power supply during extended low sunlight periods. The proposed energy effcient system can save the estimated \textbf{annual energy savings from the National Grid:}\\
Minimum: $3.46\, \text{Wh} \times 365 = \frac{1262.9}{1000} = 1.26\, \text{kWh}$ per year \\
Maximum: $34.824\, \text{Wh} \times 365 = \frac{12710.76}{1000} = 12.71\, \text{kWh}$ per year.

One of the advantages of using solar power is to reduce grid dependency. By calculating annual energy savings we can assess the long term financial and environmental benefit of the system. By generating its own power the system reduces electricity cost and contributes to a more sustainable energy solution. Over time this results in big energy savings and less reliance on conventional power sources. \\

\section{Limitations and Future Scopes}
\label{Limitations}
\subsection{Limitations}
We did not train the model to identify real bus-stops in the present version of our system. Rather as a bus-stop detection prototype, we use stop-signs from the pre-trained YOLOv4-Tiny datasets. Although this method may not accurately represent real-world circumstances, it did enable us to test the object detection capabilities. Additionally, our RFID system of the project only counts the number of seats that are available in the bus. Furthermore, the software currently displays information using dummy data. As a result, passengers can not obtain up-to-date information on bus destinations. The practical applicability of the system is limited by its lack of real-time interaction with actual buses.\\
\subsection{Future Scopes}
We will build and train our own datasets for bus-stops. This will increase system dependability and improve the accuracy of identifying real bus-stop. We intend to replace the YOLOv4-Tiny with a more potent CNN model if we are able to get better hardware resources. A robust model will enable faster processing and better accuracy, increasing the system reliability. We hope to include RFID technology with a real time fare collection system in future. RFID cards will be used by passengers to pay for their fares, which will help the transportation revenue management. We intend to give passengers real-time location information by enhancing the software. Through the software, users will be able to view their location, track their trip expenses and obtain fare related information. Also by the software, the ability for passengers to view bus arrival information in real time at their stop will be another significant enhancement. By cutting down on waiting times and uncertainty at bus-stops, this functionality will improve the user experience.\\

\section{Conclusions}
\label{end}

Although public transit is essential for urban mobility, daily commutes are nevertheless impacted by inefficiencies such erratic stopping, a lack of real-time passenger information and inadequate safety precautions. These difficulties lead to poor user experience, overall safety hazards and traffic congestion. Furthermore, the lack of sustainable energy solutions in transportation infrastructure increases reliance on traditional power sources and decreases efficiency. Adopting technologies like automated fare collecting, smart bus-stop, real-time tracking and sustainable energy solutions are crucial to build a better, safer and more effective public transportation system. Public transportation can be made more accessible by utilizing developments in computer vision, IoT and renewable energy. By incorporating smart technologies that improve safety, encourage energy efficiency and offer real-time information, we hope to make a transportation system more effective, user-friendly and well-organized. With these upgrades, we hope to give passengers a smooth and dependable experience using a smart public transportation system. \\

\section{Statements and Declarations}
\subsection{\textbf{Acknowledgements}}
None.
\subsection{\textbf{Funding}}
None Declared.
\subsection{\textbf{Conflict of Interest}}
None Declared.
\subsection{\textbf{Declaration of generative AI and AI-assisted technologies in the writing process}}
None Declared.

% \bibliographystyle{elsarticle-harv}
\bibliographystyle{unsrt}
\bibliography{ref}



% %% For authoryear reference style

% % \bibitem[Lamport(1994)]{lamport94}
% %   Leslie Lamport,
% %   \textit{\LaTeX: a document preparation system},
% %   Addison Wesley, Massachusetts,
% %   2nd edition,
% %   1994.

% \bibitem[Virgilio et al., 2020]{lit1}
%     Virgilio, G., V.R., Sossa, H., Zamora, E. (2020). Vision-Based Blind Spot Warning System by Deep Neural Networks. In: Figueroa Mora, K., Anzurez Marín, J., Cerda, J., Carrasco-Ochoa, J., Martínez-Trinidad, J., Olvera-López, J. (eds) Pattern Recognition. MCPR 2020. Lecture Notes in Computer Science(), vol 12088. Springer, Cham. https://doi.org/10.1007/978-3-030-49076-8_18

% \end{thebibliography}

\end{document}